\documentclass[12pt,prl,preprint,showpacs,showkeys]{revtex4}
\usepackage{amsfonts}

\usepackage{graphicx}
\usepackage{dcolumn}
\usepackage{bm}
\include{biblio}

\begin{document}

\title{Micro-joule sub-10-fs VUV pulse generation by MW pump pulse using highly efficient chirped-four-wave mixing in hollow-core photonic crystal fibers}
\author{Song-Jin Im}
\affiliation{Department of Physics, Kim Il Sung University, Daesong District, Pyongyang, DPR Korea}

\begin{abstract}
We theoretically study chirped four-wave mixing for VUV pulse generation in hollow-core photonic crystal fibers. 
We predict the generation of sub-10-fs VUV pulses with energy of up to hundreds of $\mu$J by broad-band chirped idler pulses 
at 830 nm and MW pump pulses with narrow-band at 277nm. 
MW pump could be desirable to reduce the complexity of the laser system or use a high repetition rate-laser system. 
The energy conversion efficiency from pump pulse to VUV pulse reaches to 30\%. 
This generation can be realized in kagome-lattice hollow-core PCF filled with noble gas of high pressure with core-diameter 
less than 40$\mu$m which would enable technically simple or highly efficient coupling to fundamental mode of the fiber.

\end{abstract}



\maketitle

Photonic crystal fibers (PCF) had a great impact on the field of nonlinear optics and led to numerous advances in other areas of physics \cite%
{russell_science_2003}. The hollow-core PCF (HC-PCF) \cite%
{cregan_science_1999}, in which light is guided in the hollow core via the bandgap effect in the cladding consisting of a two-dimensional triangular-lattice array of air holes, offer low-loss transmission with loss coefficients in the range of 1dB/km enabling the study of diverse nonlinear effects with greatly increased interaction lengths at reduced power levels in gases  \cite%
{couny_prl_2007,gosh_prl_2005,benabid_optexp_2005} or on the other hand the guidance of high-peak-power pulses \cite%
{ouzounov_science_2003}. However, the main drawback of these fibers is their intrinsically narrow transmission bandwidth determined by the bandgaps, which excludes its implementation in a large number of applications in ultrafast nonlinear optics requiring broadband guidance or guidance in the visible and ultraviolet (UV).

An alternative HC-PCF design replaces the triangular-lattice cladding with a kagome-lattice \cite%
{benabid_science_2002,couny_optlet_2006,couny_science_2007,im_optexp_2009,im_optexp_2010,im_pra_2010,joly_prl_2011,mak_optexp_2013}. The photonic guidance of this fiber is not based on a bandgap but on the inhibited coupling between the core and cladding modes. Due to this guiding mechanism kagome-lattice HC-PCFs exhibit broad transmission regions with a loss lower than 1dB/m covering the spectral range from the infrared up to the vacuum ultraviolet (VUV). As theoretically shown in \cite%
{im_optexp_2009}, these fibers exhibit controlled anomalous dispersion for UV or visible wavelengths for diameters in the range of 10 µm to 80 µm. The controllable dispersion property as well as the ultrabroadband guidance property of these fibers has been used for the generation of high-power soliton-induced supercontinuum and wavelength-tunable ultrashort VUV pulses \cite%
{im_optexp_2010,joly_prl_2011,mak_optexp_2013} and the soliton delivery of few-cycle optical gigawatt pulses \cite%
{im_pra_2010}.

Ultrashort pulse sources in the VUV spectral range are essential tools in physics, chemistry, biology and material sciences requiring further progress in the development of VUV-fs-pulse generation methods. Note that up to now no standard method for the generation of ultrashort pulses in the VUV range exist, and only relatively modest results compared with the progress in the near-infrared spectral range has been achieved, remarkable here is e.g. the generation of 11-fs pulses at 162nm with 4nJ energy \cite%
{kosma_optlet_2008}. 
A method for VUV-fs-pulse generation with a possible much higher pulse energy is the use of non-resonant four wave mixing (FWM) in hollow waveguides filled with a noble gas. Generation of 8-fs-pulses with 1$\mu$J at 270nm were reported with this method by pumping with the fundamental and second harmonic of a Ti:sapphire laser \cite%
{drufee_optlet_1999}. 
This method has been extended into the VUV range generating 600nJ, 160-fs pulses at 160nm using the fundamental and third harmonic of a Ti:sapphire laser as idler and pump, respectively: $\omega_{S}$=2$\omega_{P}$-$\omega_{I}$ with $\omega_{P}$=3$\omega_{I}$ and $\lambda_{I}$=830nm \cite%
{tzankov_optexp_2007}. 
One numerical study \cite%
{babushkin_optlet_2008}
showed the potential of this method for sub-5-fs VUV pulse generation and predicted the generation of 2.5-fs pulses at 160nm by 3-fs, 800-nm idler and narrow-band pump pulses at 267nm. However, in this case the VUV pulse energy is limited to the nJ range. 
An elemental option for high efficiency and high energy VUV fs pulse generation is chirped four-wave mixing [18] in hollow waveguides. In \cite%
{babushkin_optexp-2008}, 
the generation of 7fs with 200-$\mu$J energy by chirped four-wave mixing with broadband stretched, positively chirped near-infrared idler pulses and narrow-band UV pump pulses in hollow waveguides is predicted. But a standard hollow waveguide has instinctively high leaky loss which makes the hollow waveguide with core-diameters less than 100$\mu$m not applicable. For a larger core-diameter the phase-matching pressure should be lower than 30Torr resulting in a lower gain per unit length. Pump pulses with GW peak power are necessarily required in en effective nonlinear process through hollow waveguides because of its small figure of merit as well as the low phase-matching pressure. However, MW pumping could be desirable in practice to reduce the complexity of the laser system or use a high-power diode-pumped multi-MHz laser operating at multi-megahertz instead of kilohertz repetition rates that increases the signal-to noise ratio and reduces the time required for many measurements.

As will be shown, the possibility to control dispersion in the visible and UV range combined with moderate loss and broadband transmission in kagome-lattice HC-PCF is of great interest for applications in ultrafast nonlinear optics demanding phase-matching conditions, remarkably here for the VUV pulse generations using four-wave mixing. In this paper, we theoretically study chirped four-wave mixing for VUV pulse generation in kagome-lattice HC-PCFs. We predict the generation of sub-10-fs VUV pulses with energy of up to hundreds of $\mu$J by broad-band chirped idler pulses at 830 nm and MW pump pulses with narrow band at 277nm. The energy conversion efficiency from pump pulse to VUV pulse reaches to 30\%. This generation can be realized in kagome-lattice HC-PCF filled with noble gas of high pressure with core-diameter less than 40$\mu$m which would enable technically simple or highly efficient coupling to fundamental mode of the fiber. The kagome-lattice HC-PCF has such strut-resonant wavelengths beyond the wavelengths of pump pulse and idler pulse. We would like to note that the kind of HC-PCF with parameters considered here is practically producible and have ever been already fabricated in Bath University.

\begin{figure}
\includegraphics[width=0.7\textwidth]{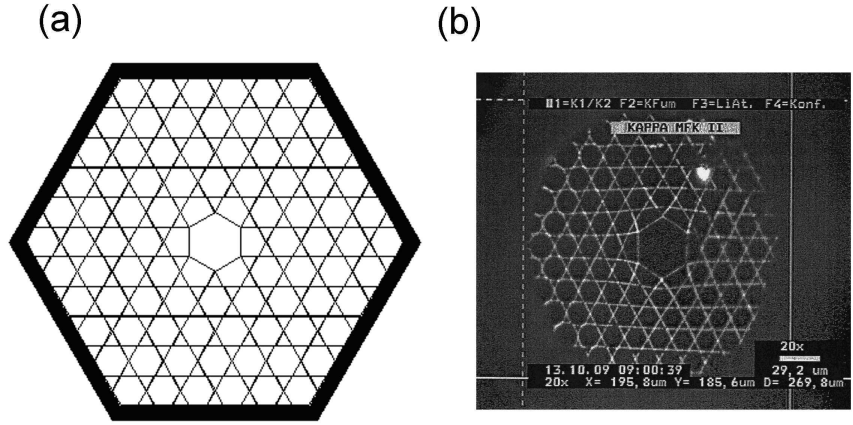}
\caption{(Color online) Cross-section of the model (a) and manufactured kagome-lattice HC-PCF (b). In (b), the photographs were taken in Max-Born Institute. The kagome-lattice HC-PCF has been fabricated in Bath University.}
\label{fig:1_Kagome}
\end{figure}

The cross-section of the studied model of kagome-lattice HC-PCF is presented in Fig.\ref{fig:1_Kagome}(a), in which a hollow core filled with a noble gas is surrounded by a kagome-lattice cladding and a bulk fused silica outer region. The dispersion of fused silica as well as that of argon was described by the Sellmeyer formula for the corresponding dielectric function. In Fig.\ref{fig:1_Kagome}(b) photographs of the cross-section of the manufactured kagome-lattice HC-PCF are shown. The kagome-lattice HC-PCF has been fabricated in Bath University and the photographs were taken in Max-Born Institute. We want to note that in this paper the kagome-lattice HC-PCF with the similar parameters as in the fabricated one [Fig.\ref{fig:1_Kagome}(b)] is considered. 

\begin{figure}
\includegraphics[width=0.7\textwidth]{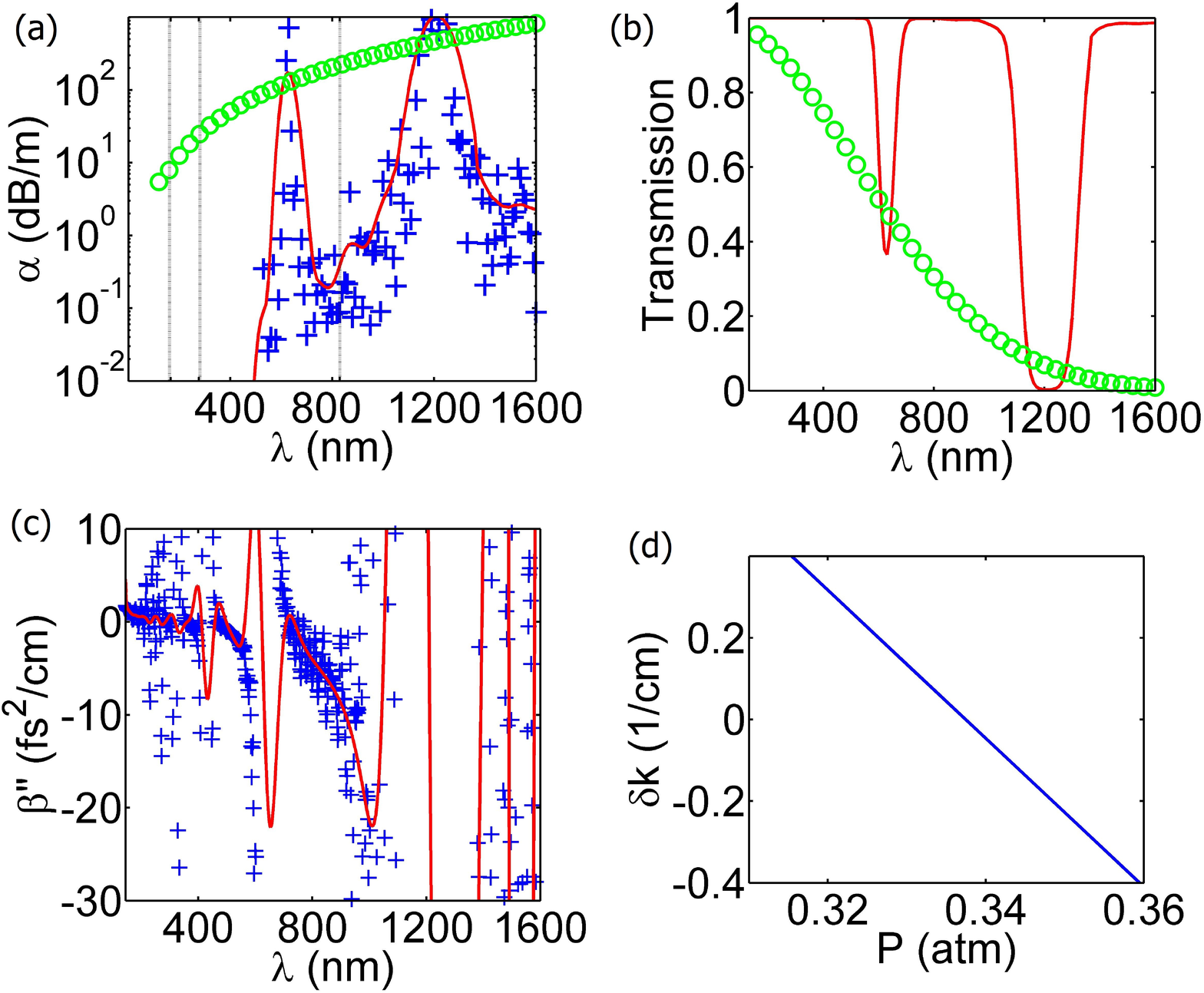}
\caption{(Color online) Loss (a), transmission through 2.5cm (b), group velocity dispersion (GVD) (c) and wave number mismatch (d) of kagome-lattice HC-PCF. In (a), (b) and (c), a kagome-lattice HC-PCF with a lattice pitch of 24$\mu$m and a strut thickness of 0.6$\mu$m filled with argon at 0.337atm is considered. In (a), the blue crosses represent the direct numerical simulations, the red solid curve is the result after averaging over inhomogeneities and the green circles are the loss of a hollow silica waveguide with the same core diameter. In (b), the red solid represents the transmission through 2.5cm of the kagome-lattice HC-PCF and the green circles are the one of hollow silica waveguide with the same core diameter. In (c), the blue crosses represent the direct numerical simulations and the red solid curve is the averaged results. In (d), the dependence of the wave number mismatch $\delta$k=2$\beta$($\omega_{P}$)-$\beta$($\omega_{I}$)-$\beta$($\omega_{S}$) on the gas pressure of the fiber is presented,where $\omega_{S}$=2$\omega_{P}$-$\omega_{I}$ with $\omega_{P}$=3$\omega_{I}$ and $\lambda_{I}$=830nm.}
\label{fig:2_Dispersion}
\end{figure}

For the calculation of the propagation constant $\beta$($\omega$) and $\alpha$($\omega$) of this waveguide the finite-element Maxwell solver JCMwave was utilized \cite%
{im_optexp_2009,im_optexp_2010,im_pra_2010}. In Fig.\ref{fig:2_Dispersion}(a) the loss coefficient is presented which is a few orders of magnitude lower than the one of a hollow silica waveguide with the same core diameter, except around the wavelengths of 1200 nm and 600 nm which coincide with the strut resonances \cite%
{couny_science_2007,im_optexp_2009}. Remarkably here it has a magnitude of below 1dB/m at around 830nm as well as 277nm and 166nm corresponding to the idler, pump and signal wavelength in the four-wave mixing as will be shown latter. This low loss is mainly influenced by the strut thickness in the kagome-lattice cladding and only weakly depends on the core diameter. We note that the guiding range of the kagome-lattice HC-PCF in the UV/VUV range is limited only by the loss of argon, while the intrinsic silica loss in the 120-200nm range does not lead to high waveguide loss \cite%
{im_optexp_2010}. In Fig.\ref{fig:2_Dispersion}(b) the transmissions through 2.5cm of the kagome-lattice HC-PCF and the hollow silica waveguide are shown.

Figure \ref{fig:2_Dispersion}(c) demonstrates the possibility to achieve comparatively small group velocity dispersion (GVD) at considered wavelengths which is not achievable in conventional photonic bandgap fibers. In a real waveguide, there are longitudinal variations of the structure parameters due to manufacturing imperfections, leading to fast longitudinal variation of the propagation constant, which however will be smoothed out during propagation. This smoothing can be also performed in the frequency domain, since the position of the spikes in the loss and dispersion curves scales correspondingly with the varying structure parameters. We assume a 5\% variation depth of the inhomogeneity and consider an averaged loss and GVD, as depicted by the red solid curves in Fig.\ref{fig:2_Dispersion}(a), (c). In Fig.\ref{fig:2_Dispersion}(d) by the blue solid the dependence of the wavelength number mismatch $\delta$k=2$\beta$($\omega_{P}$)-$\beta$($\omega_{I}$)-$\beta$($\omega_{S}$) on the gas pressure is shown, where $\omega_{S}$=2$\omega_{P}$-$\omega_{I}$ with $\omega_{P}$=3$\omega_{I}$ and $\lambda_{I}$=830nm. One can see that the phase-matching pressure is between 0.33 and 0.34atm and in this region of pressure the wavelength number mismatch is below 0.2/cm.

For the numerical simulations we use a generalized version of the 
propagation equation for the electric field strength $E$ of forward-going 
waves \cite%
{husakou_prl_2001,im_optexp_2010}.

\begin{eqnarray}
\frac{\partial E(z,\omega)}{\partial z}=i\left(\beta(\omega)-\frac{\omega}{c}%
\right)E(z, \omega)-\frac{\alpha(\omega)}{2}E(z,\omega)+  \nonumber \\
\frac{i\omega^2}{2c^2\epsilon_0\beta_{j}(\omega)}P_{NL}(z,\omega)
\label{eq:one}
\end{eqnarray}
where $P_{NL}$ describes the nonlinear Kerr polarization as well as the
photoionization-induced nonlinear absorption and phase modulation by the plasma,
and $z$ is the axial coordinate. This equation does not rely on the slowly-varying envelope approximation, includes dispersion to all orders, and can be used for the
description of extremely broad spectra. Since the transfer to higher-order
transfer modes is small, we consider only the fundamental linearly-polarized
HE$_{11}$-like mode. The nonlinear refractive index of argon is $n_{2}=1\times10^{-19}$ cm$^{2}$/W/atm.

\begin{figure}
\includegraphics[width=0.7\textwidth]{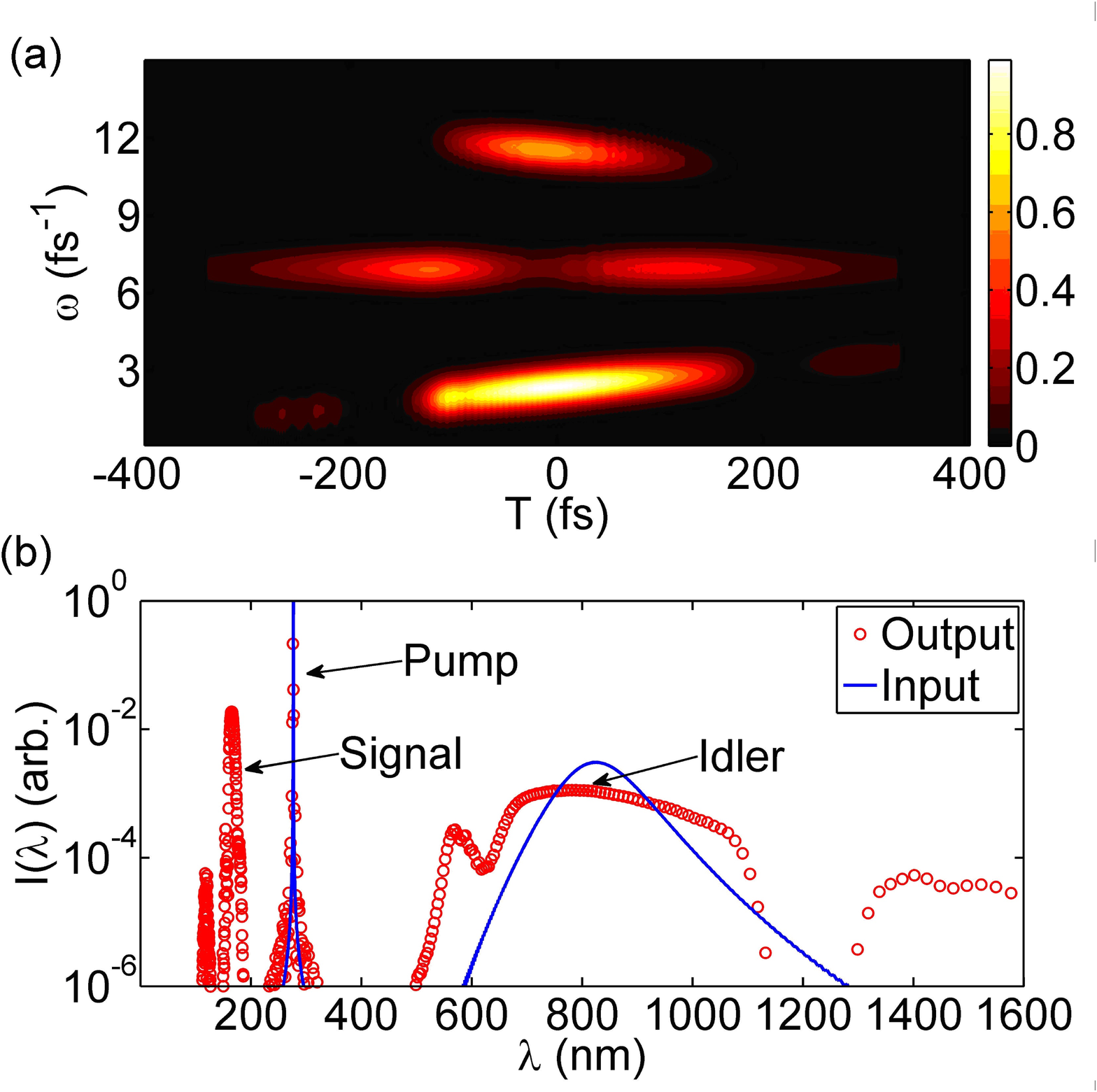}
\caption{(Color online) High-power VUV chirped pulse generation using chirped four-wave mixing in a kagome-lattice HC-PCF filled with argon gas at the phase-matching pressure of 0.337atm. Here the input peak intensity is I$_{P}$=I$_{I}$=80TW/cm$^{2}$, the input pulse duration is 300fs and the input pump frequency is $\omega_{P}$=3$\omega_{I}$ with $\lambda_{I}$=830nm. The input idler broadband pulse stretched from 6fs to the pump pulse duration by propagation through a piece of dispersive MgF$_{2}$ glass. The spectrogram projected onto the temporal intensity (a) and spectrum (b) for the pulses after propagating through 2.5cm of the fiber are shown. In (b) the red circles represent the output spectrum and blue solid is the input spectrum. The parameters of the fiber are as in Fig.\ref{fig:2_Dispersion}.}
\label{fig:3_Chirp}
\end{figure}

In Fig.\ref{fig:3_Chirp} the spectrogram projected onto the temporal intensity (a) and spectrum (b) for the pulses after propagating through 2.5cm of the kagome-lattice HC-PCF with the same parameters as in Fig.\ref{fig:2_Dispersion} filled with argon gas at the phase-matching pressure of 0.337atm. Here the input peak intensity is I$_{P}$=I$_{I}$=80TW/cm$^{2}$, the input pulse duration is 300fs and the input pump frequency is $\omega_{P}$=3$\omega_{I}$ with $\lambda_{I}$=830nm. The input idler broadband pulse stretched from 6fs to the pump pulse duration by propagation through a piece of dispersive MgF$_{2}$ glass. In (a) we can see three bright strips around 2.27fs$^{-1}$, 6.81fs$^{-1}$ and 11.35fs$^{-1}$ respectively corresponding to the idler frequency $\omega_{I}$, pump frequency $\omega_{P}$ and signal frequency $\omega_{S}$ of the four-wave mixing process. The idler strip has an upward slop and the signal strip a downward slop showing that the generated signal pulse has a negative chirp opposite to the chirp of the idler pulse. In (b) the output spectrum is presented. Here the peak around 166nm shows the signal wavelength generation by the four-wave mixing. The smaller peak near the signal wavelength peak is due to a cross-phase modulation. The spectral broadening of the idler is also due to a cross-phase modulation and the gap around 1200nm in the idle spectrum is caused by the high loss of the fiber in those wavelengths as shown in Fig.\ref{fig:2_Dispersion}(a), (b).

\begin{figure}
\includegraphics[width=0.7\textwidth]{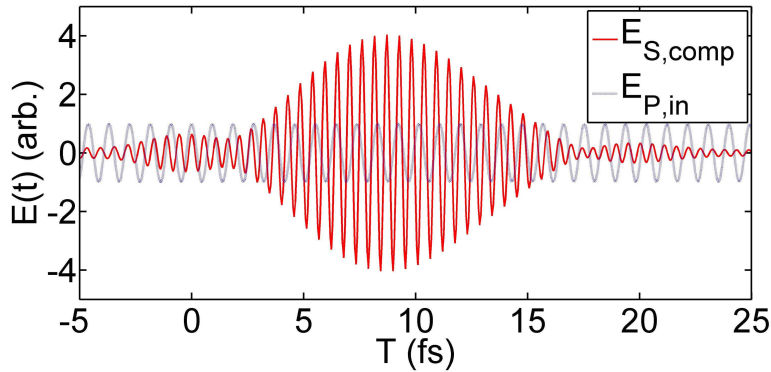}
\caption{(Color online) Electric field strength of sub-10-fs VUV pulse after compression. This pulse is obtained by compressing the signal pulse as in Fig. 3 by compensating its negative chirp in propagating through a piece of dispersive MgF$_{2}$ glass.  The red solid represents the electric filed strength of the sub-10-fs VUV pulse and the blue dotted is the one of the input pump pulse shown here for a comparison.}
\label{fig:4_Pulse}
\end{figure}

Fig.\ref{fig:4_Pulse} shows the sub-10-fs VUV pulse obtained by compressing the signal pulse as in Fig.\ref{fig:3_Chirp} compensating its negative chirp in propagating through a piece of dispersive MgF$_{2}$ glass. The peak power of the VUV pulse is about 42 times of the one of the input pump pulse and the pulse duration is about 6fs. The energy conversion efficiency from the pump pulse to the VUV pulse is about 30\% and the VUV pulse energy is about 80$\mu$J.

\begin{figure}
\includegraphics[width=0.7\textwidth]{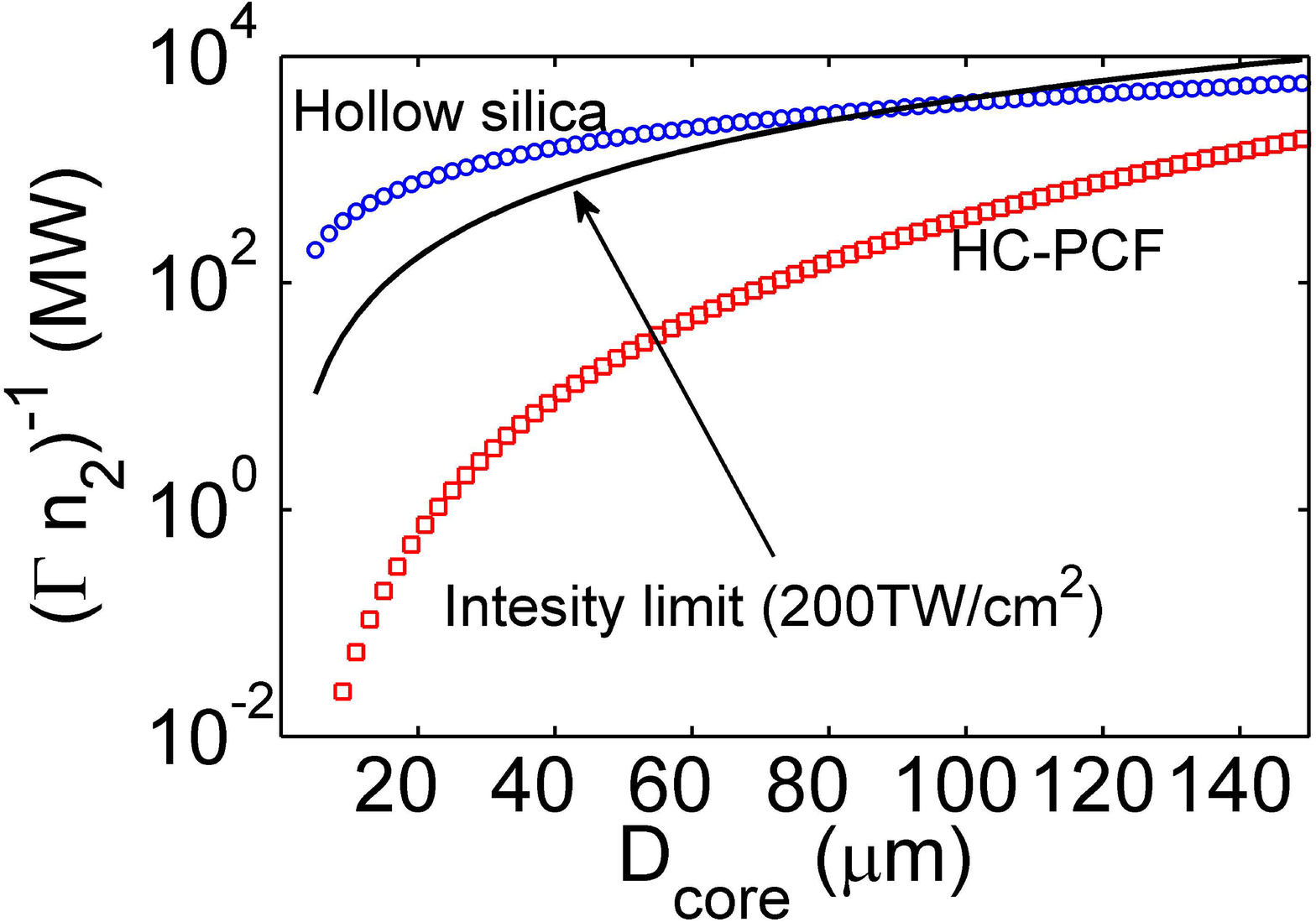}
\caption{(Color online) Inverse power nonlinearities $(\Gamma n_{2})^{-1}$ of kagome-lattice HC-PCF (blue circles) and hollow silica waveguide (red squares) filled with argon gas at proper phase-matching pressures according to the core diameters. The black solid line shows a limit by the highest peak intensity of 200TW/cm$^2$.}
\label{fig:5_Merit}
\end{figure}

Fig.\ref{fig:5_Merit} shows the inverse power nonlinearities $(\Gamma n_{2})^{-1}$ of kagome-lattice HC-PCF and hollow silica waveguide filled with argon gas at proper phase-matching pressures according to the core diameters. Here $\Gamma=2\pi L_{loss}/\lambda A_{eff}$ is the figure of merit defined as similar as in \cite%
{skibina_natphot_2008}, where $L_{loss}$ is 3dB attenuation length describing how far does the light travel before it is absorbed and $A_{eff}$ is the effective mode area.  This coefficient $\Gamma$ describes the contribution of a fiber geometry to the nonlinear phase shift $\varphi_{nl}=n_{2}\Gamma P_{0}$ that a pulse with peak power $P_{0}$ experiences during guiding in the given geometry with a nonlinear refractive index $n_{2}$ of the gas. The inverse power nonlinearities $(\Gamma n_{2})^{-1}$ describe the lowest peak power needed to take a significant nonlinear effect in a waveguide. Hollow waveguides provide losses inversely proportional to the third power of their core-diameters and a tolerable level of loss only for diameters larger than 100$\mu$m that is illustrated in the Fig.\ref{fig:5_Merit} by the thing that the inverse power nonlinearities for the diameters smaller than 100$\mu$m is above the intensity limit curve. For diameters more than 100$\mu$m at least GW peak power is needed to obtain a significant nonlinear effect. However, the loss of kagome-lattice HC-PCF is mainly determined by the cladding structure such as the strut thickness and not greatly influenced by the core-diameter and remain a few orders of magnitude below that of hollow silica waveguide with the same core-size. Therefore as shown in Fig.\ref{fig:5_Merit} for the kagome-lattice HC-PCF the needed peak power is of the orders of MW and even down to 0.01MW.

We predict micro-joule sub-10-fs VUV pulse generation by MW pump pulse using highly efficient chirped four-wave mixing in kagome-lattice HC-PCF. Low dispersion of kagome-lattice HC-PCF gives the possibility to satisfy the phase-matching condition through controlling the gas-pressure. The large figure of merit and small inverse power nonlinearity enable a MW pump-based highly efficient chirped four-wave mixing for VUV pulse generation. Using the model of kagome-lattice HC-PCF with the similar parameters as of the manufactured one in Bath University [Fig.\ref{fig:1_Kagome}(b)] we numerically obtained the 80-$\mu$J, 6-fs VUV pulse by the 240MW pump pulse. We can use a lower pump power down to 10MW in this fiber as shown in Fig.\ref{fig:5_Merit}. Moreover decreasing the core-diameter enables a lower pump power keeping a high intensity because the loss of a kagome-lattice HC-PCF is not greatly influenced by the core-diameter. MW pump could be desirable to reduce the complexity of the laser system or use the high repetition rate-laser system that increases the signal-to noise ratio and reduces the time required for many measurements.

We like to acknowledge valuable advices and supports of Dr. Herrmann and Dr. Steinmeyer in Max-Born Institute.

\end{document}